\documentstyle[12pt]{article}
\begin{document}
\title{New Mechanism For Mass Generation of Gauge Field
\thanks{This work is suppoeted by China National Natural 
Science Foundation and China Postdoctorial Science
Foundation. }}
\author{{Ning Wu}
\thanks{email address: wuning@heli.ihep.ac.cn}
\\
\\
{\small Institute of High Energy Physics, P.O.Box 918-1, 
Beijing 100039, P.R.China}
\thanks{mailing address}}
\maketitle
\vskip 0.8in

~~

\vskip 0.8in

\begin{abstract}
A new mechanism for mass generation of gauge field is discussed
in this paper. By introducing two sets of gauge fields 
and making the variations of these two sets of gauge fields 
compensate each other under local gauge transformations, the mass term 
of gauge fields is introduced into the Lagrangian without violating the
local gauge symmetry of the Lagrangian. This model is a renormalizable
quantum model. 
\\

\end{abstract}

\newpage

\Roman{section}

\section{Introduction}

Now, it is generally  believed that four
kinds of fundamental interactions in Nature are all gauge interactions
and can be described by gauge field theory.
From theoretical point of view, the principle of local gauge invariance
plays a fundamental role in particle's interaction theory\cite{1}.
According to experimental results, 
some gauge bosons are massive \cite{pdg}.
A usual way to make gauge field obtain non-zero mass
is to use spontaneously symmetry
breaking and Higgs mechanism \cite{5,6,7,8,9,10,11,12},
which is well-known in constructing the standard model \cite{2,3,4}.
But the Higgs mechanism is not the only mechanism that make
gauge field massive.  In this paper, another mechanism
for the mass generation of gauge field is proposed.
By introducing two sets of gauge fields, we will
introduce the mass term of gauge fields without violating the
local gauge symmetry of the Lagrangian. 
Because the Lagrangian
has strict local gauge symmetry, the model is renormalizable\cite{15}.
\\

\section{The Lagrangian of The Model}

Suppose that the gauge symmetry of the theory is $SU(N)$ group,   
$\psi (x)$ is a N-component vector in the fundamental representative 
space of $SU(N)$ group and the representative matrices of the
generators of $SU(N)$ group are denoted by
$T_i ~(i=1,2, \ldots, N^2-1)$. They are Hermit and traceless.
They satisfy:
$$
\lbrack T_i ~ ,~ T_j \rbrack = i f_{ijk} T_k,~~~~
Tr( T_i  T_j ) = \delta_{ij} K,
\eqno{(2.1)} 
$$
where $f_{ijk}$ are structure constants of $SU(N)$ group, K is 
a constant which is independent of indices $i$ and $j$ but depends 
on the representation of the group. The representative matrix 
of a general element of the $SU(N)$ group is expressed as:
$$
U=e^{-i \alpha ^i T_i}
\eqno{(2.2)} 
$$
with $\alpha ^i$ the real group parameters. 
In global gauge transformations, all $\alpha ^i$ are independent of
space-time coordinates, while in local gauge transformations, 
$\alpha ^i$ are functions of space-time coordinates. 
$U$ is a unitary $N \times N$ matrix.
\\

In order to introduce the mass term of gauge fields 
without violating local gauge symmetry, 
two kinds of gauge fields $A_{\mu}(x)$ and $B_{\mu}(x)$ are needed. 
$A_{\mu}(x)$ and $B_{\mu}(x)$ are vectors in the canonical 
representative space of $SU(N)$ group. They can be expressed as linear 
combinations of generators :
$$
A_{\mu}(x) = A_{\mu} ^i (x) T_i
\eqno{(2.3a)} 
$$
$$
B_{\mu}(x) = B_{\mu} ^i (x) T_i.
\eqno{(2.3b)}
$$
where $A_{\mu}^i (x)$ and $B_{\mu}^i (x)$ are component fields of 
gauge fields $A_{\mu}(x)$ and $B_{\mu}(x)$ respectively. \\

Corresponding to two kinds of gauge fields, there are two 
kinds of gauge  covariant derivatives in the theory:
$$
D_{\mu} = \partial _{\mu} - ig A_{\mu}
\eqno{(2.4a)} 
$$
$$
D_{b \mu} = \partial _{\mu} + i \alpha g B_{\mu}.
\eqno{(2.4b)} 
$$
The strengths of gauge fields $A_{\mu}(x)$ and $B_{\mu}(x)$ 
are defined as 
$$
\begin{array}{ccl}
A_{\mu \nu} & = & \frac{1}{-i g} \lbrack D_{\mu} ~,~ D_{\nu} \rbrack
\\
& = & \partial _{\mu} A_{\nu} - \partial _{\nu} A_{\mu}
- i g \lbrack A_{\mu} ~,~ A_{\nu} \rbrack
\end{array}
\eqno{(2.5a)} 
$$
$$
\begin{array}{ccl}
B_{\mu \nu} &=& \frac{1}{i \alpha g} \lbrack D_{b \mu} 
~,~ D_{b \nu} \rbrack
\\
& = & \partial _{\mu} B_{\nu} - \partial _{\nu} B_{\mu}
+ i \alpha g \lbrack B_{\mu} ~,~ B_{\nu} \rbrack. 
\end{array}
\eqno{(2.5b)} 
$$
respectively. Similarly, $A_{\mu \nu}$ and $B_{\mu \nu}$ can 
also be expressed as linear combinations of generators:
$$
A_{\mu \nu} = A_{\mu \nu}^i T_i
\eqno{(2.6a)} 
$$
$$
B_{\mu \nu}= B_{\mu \nu}^i T_i. 
\eqno{(2.6b)} 
$$
Using relations (2.1) and (2.5a,b), we can obtain
$$
A_{\mu \nu}^i = \partial _{\mu} A_{\nu}^i - \partial _{\nu} A_{\mu}^i
+g f^{ijk} A_{\mu}^j    A_{\nu}^k
\eqno{(2.7a)} 
$$
$$
B_{\mu \nu}^i = \partial _{\mu} B_{\nu}^i - \partial _{\nu} B_{\mu}^i
- \alpha g f^{ijk} B_{\mu}^j    B_{\nu}^k .
\eqno{(2.7b)} 
$$
\\

 The Lagrangian density of the model is
$$
\begin{array}{ccl}
\cal L &= &- \overline{\psi}(\gamma ^{\mu} D_{\mu} +m) \psi 
-\frac{1}{4K} Tr( A^{\mu \nu} A_{\mu \nu} )
-\frac{1}{4K} Tr( B^{\mu \nu} B_{\mu \nu} ) \\
&&-\frac{\mu ^2}{2K ( 1+ \alpha ^2)} 
Tr \left \lbrack (A^{\mu}+\alpha B^{\mu})( A_{\mu}+\alpha B_{\mu} ) 
\right \rbrack
\end{array}
\eqno{(2.8)} 
$$
where $\alpha$ is a constant. In this paper, the space-time  
metric is selected as $ \eta _{\mu \nu} = diag (-1,1,1,1)$, 
$(\mu ,\nu =0,1,2,3)$. According to relation (2.1), the above 
Lagrangian density  ${\cal L}$ can be rewritten as:
$$
\begin{array}{ccl}
\cal L &= &- \overline{\psi} \lbrack \gamma ^{\mu} ( \partial _{\mu}
 - i g A^i_{\mu} T_i) +m \rbrack \psi 
-\frac{1}{4}  A^{i \mu \nu} A^i_{\mu \nu} 
-\frac{1}{4}  B^{i \mu \nu} B^i_{\mu \nu} \\
&&-\frac{\mu ^2}{2 ( 1+ \alpha ^2)} 
(A^{i \mu}+\alpha B^{i \mu})( A^i_{\mu}+\alpha B^i_{\mu} ) .
\end{array}
\eqno{(2.9)} 
$$
An obvious characteristic of the above Lagrangian is that the mass 
term of the gauge fields is introduced into the Lagrangian and this 
term does not affect the symmetry of the Lagrangian. We will prove
that the above Lagrangian has strict local gauge symmetry in 
the chapter 4.
\\

Because both vector fields $A_{\mu}$ and $B_{\mu}$ are
standard gauge fields,  this model is a kind of 
gauge field model which describes
gauge interactions between gauge fields and matter fields.
\\

\section{Global Gauge Symmetry and Conserved Charges}

~~~~ Now, let's  discuss the gauge symmetry of the Lagrangian 
density ${\cal L}$. First, we will discuss the global gauge 
symmetry and the corresponding conserved charges. In global 
gauge transformation, the matter field $\psi$ transforms as:
$$
\psi \longrightarrow \psi ' = U \psi ,
\eqno{(3.1)} 
$$
where $U$ is independent 
of space-time coordinates. That is 
$$
\partial _{\mu} U = 0. 
\eqno{(3.2)} 
$$
The corresponding global gauge transformations of gauge 
fields $A_{\mu}$ and $B_{\mu}$ are
$$
A_{\mu} \longrightarrow U A_{\mu} U^{\dag}
\eqno{(3.3a)} 
$$
$$
B_{\mu} \longrightarrow U B_{\mu} U^{\dag}
\eqno{(3.3b)} 
$$
respectively. It is easy to prove that 
$$
D_{\mu} \longrightarrow U D_{\mu} U^{\dag}
\eqno{(3.4a)} 
$$
$$
D_{b \mu} \longrightarrow U D_{b \mu} U^{\dag}
\eqno{(3.4b)} 
$$
$$
A_{\mu \nu} \longrightarrow U A_{\mu \nu} U^{\dag}
\eqno{(3.5a)} 
$$
$$
B_{\mu \nu} \longrightarrow U B_{\mu \nu} U^{\dag}
\eqno{(3.5b)} 
$$
Using all the above transformation relations, it can be strictly 
proved that all terms in eq(2.8) are gauge invariant. So, the 
whole Lagrangian density has global gauge symmetry.  \\

 Let $\alpha ^i$ in eq(2.2) be the first order infinitesimal 
parameters, then, in the first order approximation,  the 
transformation matrix $U$ can be rewritten as:
$$
U \approx 1 - i \alpha ^i T^i .
\eqno{(3.6)} 
$$
The first order infinitesimal variations of fields 
$\psi, ~\overline{\psi}, ~A_{\mu}$ and $B_{\mu}$ are
$$
\delta \psi = - i \alpha ^i T^i \psi
\eqno{(3.7a)} 
$$
$$
\delta  \overline {\psi} =  i \alpha ^i \overline{\psi} T^i
\eqno{(3.7b)} 
$$
$$
\delta A_{\mu} =  \alpha ^i f^{ijk} A^j_{\mu} T^k 
\eqno{(3.8a)} 
$$
$$
\delta B_{\mu} =  \alpha ^i f^{ijk} B^j_{\mu} T^k 
\eqno{(3.8b)} 
$$
respectively. From eqs(3.8a,b) and eqs(2.3a,b), we can obtain
$$
\delta A_{\mu}^k =  \alpha ^i f^{ijk} A^j_{\mu}
\eqno{(3.9a)} 
$$
$$
\delta B_{\mu}^k =  \alpha ^i f^{ijk} B^j_{\mu}.
\eqno{(3.9b)} 
$$
\\

 The first order variation of the Lagrangian density is 
$$
\begin{array}{ccl}
\delta {\cal L} & = &
\partial _{\mu} \left (
\frac{ \partial {\cal L}}{\partial \partial _{\mu} \psi} \delta \psi 
+  \delta \overline{\psi}  \frac{ \partial {\cal L}}
{\partial \partial _{\mu} \overline{\psi}}
+  \frac{ \partial {\cal L}}{\partial \partial 
_{\mu} A_{\nu}^k } \delta A_{\nu}^k
+  \frac{ \partial {\cal L}}{\partial \partial 
_{\mu} B_{\nu}^k } \delta B_{\nu}^k
\right )
\\
& = &
\alpha ^i \partial ^{\mu} J^i_{\mu},
\end{array}
\eqno{(3.10)} 
$$
where
$$
J^i_{\mu} = 
i \overline{\psi} \gamma _{\mu} T^i \psi
- f^{ijk} A^{j \nu} A^k _{\mu \nu}
- f^{ijk} B^{j \nu} B^k _{\mu \nu}.
\eqno{(3.11)} 
$$
The conserved current can also be written as
$$
\begin{array}{ccl}
J_{\mu} & = & 
i \overline{\psi} \gamma _{\mu} T^i \psi T^i
+ i \lbrack  A^{ \nu}  ~,~  A _{\mu \nu} \rbrack 
+ i \lbrack  B^{ \nu}  ~,~  B _{\mu \nu} \rbrack 
\\
& = & J^i _{\mu} T^i . 
\end{array}
\eqno{(3.12)} 
$$
Because the Lagrangian density ${\cal L}$ has global gauge 
symmetry, the variation of ${\cal L}$ under global gauge 
transformations vanishes. That is
$$
\delta {\cal L} = 0.
\eqno{(3.13)}
$$
Because $\alpha^i $ are arbitrary global parameters, from eq(3.10), 
we can obtain the following conservation equation:
$$
\partial ^{\mu} J^i_{\mu} = 0 .
\eqno{(3.14)}
$$
\\

 The corresponding conserved charges are
$$
\begin{array}{ccl}
Q^i & = &  \int d^3 x  J^{i 0}
\\
& = &  \int d^3 x (
\psi ^{\dag} T^i \psi 
+  \lbrack  A_j   ~,~  A ^{j 0} \rbrack  ^i
+  \lbrack  B_j   ~,~  B ^{j 0} \rbrack  ^i
) .
\end{array}
\eqno{(3.15)} 
$$
After quantization, $Q^i$ are  generators of gauge transformation. 
An important feature of the above relation is that 
no matter what the value of parameter 
$\alpha$ is, gauge fields $A_{\mu}$ and $B_{\mu}$ contribute the 
same terms to the conserved currents and conserved charges. \\

\section{Local Gauge Symmetry}

~~~~ If $U$ in eq(3.1) depends on space-time coordinates, the 
transformation of eq(3.1) is a local $SU(N)$ gauge transformation. 
In this case,
$$
\partial _{\mu} U \not= 0 ~,~~ \partial _{\mu} \alpha^i \not= 0
\eqno{(4.1)}
$$
The corresponding transformations of gauge fields $A_{\mu}$ 
and $B_{\mu}$ are 
$$
A_{\mu} \longrightarrow U A_{\mu} U^{\dag}
-\frac{1}{ig}U \partial _{\mu}U^{\dag}
\eqno{(4.2a)} 
$$
$$
B_{\mu} \longrightarrow U B_{\mu} U^{\dag}
+\frac{1}{i \alpha g}U \partial _{\mu}U^{\dag}
\eqno{(4.2b)} 
$$
respectively.\\

 Using above transformation relations, it is easy to prove that
$$
D_{\mu} \longrightarrow U D_{\mu} U^{\dag}
\eqno{(4.3a)} 
$$
$$
D_{b \mu} \longrightarrow U D_{b \mu} U^{\dag}.
\eqno{(4.3b)} 
$$
Therefore,
$$
A_{\mu \nu} \longrightarrow U A_{\mu \nu} U^{\dag}
\eqno{(4.4a)} 
$$
$$
B_{\mu \nu} \longrightarrow U B_{\mu \nu} U^{\dag}
\eqno{(4.4b)} 
$$
$$
D_{\mu} \psi  \longrightarrow U D_{\mu} \psi
\eqno{(4.5)} 
$$
$$
A_{\mu} + \alpha B_{\mu} \longrightarrow 
U  (A_{\mu} + \alpha B_{\mu}  )  U^{\dag}
\eqno{(4.6)} 
$$
It can be strictly proved  that the Lagrangian density 
${\cal L}$ defined by eq(2.8) is invariant under the above local
$SU(N)$ gauge transformations. Therefore the model has strict 
local gauge symmetry.  \\

An obvious characteristics of this gauge field theory is 
that two different gauge fields $A_{\mu}$ and $B_{\mu}$ which 
correspond  to one gauge symmetry are introduced into the theory. 
From eq(4.2a,b), we know that both gauge fields $A_{\mu}$ and 
$B_{\mu}$ are standard gauge fields. But, they have different 
roles in theory. It is known that, in Yang-Mills theory, gauge
field can be regarded as gauge compensatory field of matter 
fields. In other words, if there were no gauge field, though the 
Lagrangian could have global gauge symmetry, it would have no 
local gauge symmetry. In order to make the Lagrangian have local 
gauge symmetry, we must introduce gauge field and make the 
variation of gauge field under local gauge transformations
compensate the variation of the kinematical terms of matter 
fields. So, the form of local gauge transformation of gauge 
field is determined by the form of local gauge transformation of
matter fields. Similar case holds in the gauge field theory
which is discussed in this paper: gauge field $A_{\mu}$ can 
regarded as gauge compensatory field of matter fields and gauge 
field $B_{\mu}$ can be regarded as gauge compensatory field 
of gauge field $A_{\mu}$.  Therefore, in this gauge field 
theory, the form of the local gauge transformation of gauge 
field $A_{\mu}$ is determined by the form of local gauge 
transformation of matter fields, and the form of local gauge
transformation of gauge field $B_{\mu}$ is determined by the 
form of local gauge transformation of gauge field $A_{\mu}$.
And because of the compensation of gauge field $B_{\mu}$, the 
mass term of gauge field can be introduced into the Lagrangian 
without violating its local gauge symmetry.  \\

\section{The Masses of Gauge Fields}

The mass term of gauge fields can be written as:
$$
( A^{\mu} ~,~ B^{\mu} )  M  \left (
\begin{array}{c} 
A_{\mu}  \\
B_{\mu}
\end{array}
\right ) .
\eqno{(5.1)}
$$
where $M$ is the mass matrix:
$$
M  = \frac{1}{1+\alpha ^2} \left ( 
\begin{array}{cc}
\mu ^2  &  \alpha \mu^2  \\
\alpha \mu ^2  &  \alpha^2 \mu^2
\end{array}
\right ).
\eqno{(5.2)} 
$$
Generally speaking, physical particles generated from gauge 
interactions  are eigenvectors of mass matrix and the 
corresponding masses  of these particles are eigenvalues of 
mass matrix. $M$ has two eigenvalues, they are
$$
m^2_1 = \mu ^2 ~~,~~  m^2_2 = 0.
\eqno{(5.3)} 
$$
The corresponding eigenvectors are
$$
\left (
\begin{array}{c}
{\rm cos } \theta  \\  {\rm sin} \theta
\end{array}
\right)
~~ , ~~
\left (
\begin{array}{c}
- {\rm sin } \theta  \\  {\rm cos} \theta
\end{array}
\right) ,
\eqno{(5.4)}
$$
where,
$$
{\rm  cos} \theta =  \frac{1}{\sqrt{1+\alpha ^2}}
  ~~,~~  
{\rm  sin} \theta =  \frac{\alpha}{\sqrt{1+\alpha ^2}} .
\eqno{(5.5)} 
$$
\\

Define
$$
C_{\mu}={\rm cos}\theta A_{\mu}+{\rm sin}\theta B_{\mu}
\eqno{(5.6a)} 
$$
$$
F_{\mu}=-{\rm sin}\theta A_{\mu}+{\rm cos}\theta B_{\mu}.
\eqno{(5.6b)} 
$$
It is easy to see that $C_{\mu}$ and $F_{\mu}$ are eigenstates 
of mass matrix, they describe those particles generated from 
gauge interactions. The inverse transformations of (5.6a,b) are 
$$
A_{\mu}= {\rm cos}\theta C_{\mu} - {\rm sin}\theta F_{\mu}
\eqno{(5.7a)} 
$$
$$
B_{\mu}= {\rm sin}\theta C_{\mu}+{\rm cos}\theta F_{\mu}.
\eqno{(5.7b)} 
$$
Then the Lagrangian density ${\cal L}$ given by (2.9) changes into:
$$
{\cal L} = {\cal L}_0 + {\cal L}_I ,
\eqno{(5.8)}
$$
where
$$
{\cal L}_0= - \overline{\psi}(\gamma ^{\mu} \partial _{\mu} +m) \psi 
-\frac{1}{4} C^{i \mu \nu}_0 C^i_{0 \mu \nu} 
-\frac{1}{4} F^{i \mu \nu}_0 F^i_{0 \mu \nu}
-\frac{\mu ^2}{2} C^{i \mu} C^i_{\mu}.
\eqno{(5.9a)} 
$$
$$
\begin{array}{ccl}
{\cal L}_I & = & i g \overline{\psi} 
\gamma ^{\mu} ( {\rm cos}\theta C_{\mu}
 - {\rm sin}\theta F_{\mu} )  \psi  \\
&& 
- \frac{{\rm cos}2 \theta}{2 {\rm cos} \theta} 
g f^{ijk}C_0^{i \mu \nu} C^j_{\mu} C^k_{\nu}
+\frac{{\rm sin} \theta}{2 } 
g f^{ijk}F_0^{i \mu \nu} F^j_{\mu} F^k_{\nu}  \\
&&
+\frac{{\rm sin} \theta}{2 } 
g f^{ijk}F_0^{i \mu \nu} C^j_{\mu} C^k_{\nu}  
+ g {\rm sin} \theta f^{ijk} 
C_0^{i \mu \nu} C^j_{\mu} F^k_{\nu}  \\
&&
- \frac{1 - \frac{3}{4}{\rm sin}^2 2 \theta}
{4 {\rm cos}^2 \theta} g^2 
f^{ijk} f^{ilm}  C^j_{\mu} C^k_{\nu} C^{l \mu} C^{m \nu}  \\
&&
- \frac{{\rm sin}^2  \theta}{4} g^2 f^{ijk} f^{ilm} 
F^j_{\mu} F^k_{\nu} F^{l \mu} F^{m \nu}
+ g^2 {\rm tg} \theta {\rm cos} 2 \theta f^{ijk} 
f^{ilm} C^j_{\mu} C^k_{\nu} C^{l \mu} F^{m \nu}
\\
&&
- \frac{{\rm sin}^2  \theta}{2} g^2 f^{ijk} 
f^{ilm} ( C^j_{\mu} C^k_{\nu} F^{l \mu} F^{m \nu}
+ C^j_{\mu} F^k_{\nu} F^{l \mu} C^{m \nu}
+ C^j_{\mu} F^k_{\nu} C^{l \mu} F^{m \nu}) .
\end{array}
\eqno{(5.9b)} 
$$
In the above relations, we have used the following 
simplified notations:
$$
C_{0 \mu \nu}^i = \partial _{\mu} C_{\nu}^i - \partial _{\nu} C_{\mu}^i
\eqno{(5.10a)} 
$$
$$
F_{0 \mu \nu}^i = \partial _{\mu} F_{\nu}^i - \partial _{\nu} F_{\mu}^i
\eqno{(5.10b)} 
$$
\\

 From eq(5.9a), it is easy to see that the mass of  field 
$C_{\mu}$ is $\mu$ and the mass of gauge field $F_{\mu}$ 
is zero. That is
$$
m_c = \mu ~~,~~ m_F = 0 .
\eqno{(5.11)}
$$
\\

Transformations (5.7a,b) are pure algebraic operations
which do not affect the gauge symmetry of the Lagrangian.
They can be regarded as redefinitions of gauge fields. 
The local gauge symmetry of the Lagrangian is still 
strictly preserved after field transformations. In another 
words, the symmetry of the Lagrangian before transformations 
is completely the same as the symmetry of the Lagrangian
after transformations. In fact, we do not introduce any kind of
symmetry breaking in the whole paper. \\

Fields $C_{\mu}$ and $F_{\mu}$ are linear combinations of 
gauge fields $A_{\mu}$ and $B_{\mu}$, so the forms of local 
gauge transformations of fields $C_{\mu}$ and $F_{\mu}$ are 
determined by the forms of local gauge transformations of
gauge fields $A_{\mu}$ and $B_{\mu}$. 
Because $C_{\mu}$ and $F_{\mu}$ consist of gauge fields 
$A_{\mu}$ and $B_{\mu}$ and transmit gauge interactions between 
matter fields, for the sake of simplicity, we also call them 
gauge field, just as we call  $W^{\pm}$ and $Z^0$ gauge fields 
in electroweak model. Therefor,  two different kinds of 
force-transmitting vector fields exist in this gauge field 
theory: one is massive and another is massless.
\\

\section{Equation of Motion}

~~~~ The Euler-Lagrange equation of motion for
 fermion field can be deduced from eq(5.8):
$$
\lbrack \gamma ^{\mu} ( \partial _{\mu} - i g {\rm cos}\theta C_{\mu}
+ i g {\rm sin}\theta F_{\mu} ) +m \rbrack  \psi = 0 .
\eqno{(6.1)}
$$
If we deduce the Euler-Lagrange equations of motion of gauge fields from 
eq(5.8), we will obtain very complicated expressions. For the sake of 
simplicity, we deduce the equations of motion of gauge fields from eq(2.8). 
In this case, the equations of motion of gauge fields $A_{\mu}$ and 
$B_{\mu}$ are:
$$
D^{\mu} A_{\mu \nu}- \frac{\mu^2}{1+\alpha^2} ( A_{\nu} + \alpha B_{\nu}) = 
i g \overline{\psi} \gamma _{\nu} T^i \psi T^i
\eqno{(6.2a)}
$$
$$
D_b^{\mu} B_{\mu \nu}- \frac{\alpha \mu^2}{1+\alpha^2} 
( A_{\nu} + \alpha B_{\nu}) = 0 
\eqno{(6.2b)}
$$
respectively. In the above relations, we have used two simplified notations:
$$
D^{\mu} A_{\mu \nu} = \lbrack D^{\mu} ~,~ A_{\mu \nu} \rbrack
\eqno{(6.3a)}
$$
$$
D_b^{\mu} B_{\mu \nu} = \lbrack D_b^{\mu} ~,~ B_{\mu \nu} \rbrack . 
\eqno{(6.3b)}
$$
\\

 Eqs(6.2a,b) can be expressed in terms of component fields 
$A^i_{\mu}$ and $B^i_{\mu}$:
$$
\partial ^{\mu} A^i_{\mu \nu}- \frac{\mu^2}{1+\alpha^2} 
( A^i_{\nu} + \alpha B^i_{\nu}) = 
i g \overline{\psi} \gamma _{\nu} T^i \psi + g f^{ijk} 
A^j_{\mu \nu} A^{k \nu}
\eqno{(6.4a)}
$$
$$
\partial ^{\mu} B^i_{\mu \nu}- \frac{\alpha \mu^2}{1+\alpha^2} 
( A^i_{\nu} + \alpha B^i_{\nu}) 
= - \alpha  g f^{ijk} B^j_{\mu \nu} B^{k \nu}
\eqno{(6.4b)}
$$
The equations of motion for gauge fields $C_{\mu}$ and $F_{\mu}$ can be
easily obtained from eqs(6.4a,b). In other words, cos$\theta \cdot$ 
(6.4a) -- sin$\theta  \cdot$(6.4b) gives the equation of motion for 
gauge field $C_{\mu}$, and  -- sin$\theta \cdot$ (6.4a) + 
cos$\theta  \cdot$(6.4b) gives the equation of motion for gauge 
field $F_{\mu}$.  \\

 From eq(6.2a) or (6.2b), we can obtain a supplementary condition. 
Using eq(6.1), we can prove that 
$$
\lbrack D^{\lambda} ~,~ - i g \overline{\psi} 
\gamma _{\lambda} T^i \psi T^i \rbrack = 0.
\eqno{(6.5)}
$$
Let $D^{\nu}$ act on eq(6.2a) from the left, and let $D_b^{\nu}$ 
act on eq(6.2b) from the left, applying eq(5.4) and the following 
two identities: 
$$
\lbrack D^{\lambda} ~,~ \lbrack D^{\nu} 
~,~A_{\nu \lambda} \rbrack \rbrack = 0
\eqno{(6.6a)}
$$
$$
\lbrack D_b^{\lambda} ~,~ \lbrack D_b^{\nu} 
~,~B_{\nu \lambda} \rbrack \rbrack = 0 ,
\eqno{(6.6b)}
$$
we can obtain the following two equations
$$
\lbrack D^{\nu} ~,~  A_{\nu } + \alpha B_{\nu} \rbrack  = 0
\eqno{(6.7a)}
$$
$$
\lbrack D_b^{\nu} ~,~  A_{\nu } + \alpha B_{\nu} \rbrack  = 0
\eqno{(6.7b)}
$$
respectively. These two equations are essentially the same, they 
give a supplementary condition. If we expressed eqs(6.7a,b) in 
terms of component fields, these two equations will give the same 
expression:
$$
\partial ^{\nu} ( A^i_{\nu} + \alpha B^i_{\nu} ) 
+ \alpha g f^{ijk} A^j_{\nu} B^{k \nu} = 0 .
\eqno{(6.8)}
$$
\\

 When $\nu = 0$, eqs(6.4a,b) don't give dynamical equations  
for gauge fields, because they contain no time derivative 
terms. They are just constrains. Originally, gauge fields  $A^i_{\mu}$ 
and $B^i_{\mu}$ have $8(N^2-1)$ degrees of freedom, but they satisfy 
$2(N^2-1)$constrains and have $ (N^2-1)$ gauge degrees of freedom, 
therefore, gauge fields $A^i_{\mu}$  and $B^i_{\mu}$ have $5(N^2-1)$ 
independent dynamical degrees of freedom altogether. This 
result coincides with our experience: a massive vector field has 3 
independent degrees of freedom and a massless vector field has 2 
independent degrees of freedom. \\

\section{The Case That Matter Fields Are Scalar Fields}

~~~~~ In the above discussions, matter fields are spinor fields. 
Now, let's consider the case that matter fields are scalar fields. 
Suppose that there are $N$ scalar fields $\varphi _l (x) ~(l=1,2, \cdots 
N)$ which form a multiplet of matter fields:
$$
\varphi (x) =\left ( 
\begin{array}{c}
\varphi_1 (x) \\
\varphi_2 (x) \\
\vdots \\
\varphi_N (x)
\end{array}
\right ) 
\eqno{(7.1)} 
$$
All $\varphi (x)$ form a fundamental representative space of $SU(N)$ group. 
In $SU(N)$ gauge transformation, $\varphi (x)$ transforms as :
$$
\varphi (x) \longrightarrow \varphi ' (x) = U \varphi (x)
\eqno{(7.2)}
$$
\\

 The Lagrangian density is 
$$
\begin{array}{ccl}
\cal L &= &- \lbrack (\partial _{\mu} 
- i g A_{\mu} ) \varphi \rbrack ^{+}
(\partial ^{\mu} - i g A^{\mu} ) \varphi - V( \varphi ) \\
&&-\frac{1}{4K} Tr( A^{\mu \nu} A_{\mu \nu} )
-\frac{1}{4K} Tr( B^{\mu \nu} B_{\mu \nu} ) \\
&&-\frac{\mu ^2}{2K ( 1+ \alpha ^2)} 
Tr \left \lbrack (A^{\mu}+\alpha B^{\mu})( A_{\mu}+\alpha B_{\mu} ) 
\right \rbrack
\end{array}
\eqno{(7.3)} 
$$
The above Lagrangian density can be expressed in terms of 
component fields :
$$
\begin{array}{ccl}
\cal L &= &- \lbrack (\partial _{\mu} 
- i g A^i_{\mu} T_i ) \varphi \rbrack ^{+}
(\partial ^{\mu} - i g A^{i \mu} T_i ) \varphi - V( \varphi ) \\
&&-\frac{1}{4}  A^{i \mu \nu} A^i_{\mu \nu} 
-\frac{1}{4} B^{i \mu \nu} B^i_{\mu \nu}  \\
&&-\frac{\mu ^2}{2 ( 1+ \alpha ^2)} 
 (A^{i \mu}+\alpha B^{i \mu})( A^i_{\mu}+\alpha B^i_{\mu} ) 
\end{array}
\eqno{(7.4)} 
$$
The general form for $V(\varphi)$ which is renormalizable and 
gauge invariant is
$$
V(\varphi) = m^2 \varphi^{+} \varphi + \lambda (\varphi ^{+} \varphi)^2 .
\eqno{(7.5)}
$$
It is easy to prove that the Lagrangian density ${\cal L}$ defined 
by eq(7.3) has local $SU(N)$ gauge symmetry. The Euler-Lagrange 
equation of motion for scalar field $\varphi$ is:
$$
 (\partial ^{\mu} - i g A^{\mu} ) (\partial _{\mu} 
- i g A_{\mu} ) \varphi - m^2 \varphi 
-  2 \lambda \varphi ( \varphi^{+} \varphi )^2=0
\eqno{(7.6)}
$$
\\

 If $N^2-1$ scalar fields $\varphi _l (x) ~(l=1,2, \cdots N^2-1)$ 
form a multiplet of matter fields
$$
\varphi (x) = \varphi _l (x) T_l ,
\eqno{(7.7)}
$$
then, the gauge transformation of $\varphi(x)$ is
$$
\varphi (x) \longrightarrow \varphi ' (x) = U \varphi (x) U^{+} .
\eqno{(7.8)}
$$
All $\varphi (x)$ form a space of adjoint representation of $SU(N)$ 
group. In this case, the gauge covariant derivative is
$$
D_{\mu} \varphi = \partial _{\mu} \varphi 
- i g \lbrack A_{\mu} ~,~ \varphi \rbrack , 
\eqno{(7.9)}
$$
and the gauge invariant Lagrangian density ${\cal L}$ is 
$$
\begin{array}{ccl}
\cal L &= &- \frac{1}{K} 
Tr \lbrack (D^{\mu} \varphi) ^{+}(D_{\mu} \varphi) \rbrack - 
V( \varphi ) \\

&&-\frac{1}{4}  A^{i \mu \nu} A^i_{\mu \nu} 
-\frac{1}{4} B^{i \mu \nu} B^i_{\mu \nu}  \\
&&-\frac{\mu ^2}{2 ( 1+ \alpha ^2)} 
 (A^{i \mu}+\alpha B^{i \mu})( A^i_{\mu}+\alpha B^i_{\mu} ) .
\end{array}
\eqno{(7.10)} 
$$

\section{A More General Model}

In the above discussions, a gauge field model, which has strict local  
$SU(N)$ gauge symmetry and contains massive gauge bosons, is constructed. 
In the above model, only gauge field $A_{\mu}$ directly interacts with matter 
fields $\psi$ or $\varphi$, gauge field $B_{\mu}$ doesn't directly interact 
with matter fields.  But this restriction is not necessary in 
constructing the model. In this section, 
we will construct a more general gauge field 
model, in which both gauge fields interact with matter fields in the
original Lagrangian.  As an example, we 
only discuss the case that matter fields are spinor fields. The case that 
matter fields are scalar fields can be discussed similarly.  \\

 In chapter 4, we have prove that, under local gauge transformations, 
$D_{\mu}$ and $D_{b \mu}$ transform covariantly. It is easy to prove 
that ${\rm cos}^2 \phi D_{\mu} + {\rm sin}^2 \phi D_{b \mu}$ is the 
most general gauge covariant derivative which transforms covariantly 
under local $SU(N)$ gauge transformations
$$
{\rm cos}^2 \phi D_{\mu} + {\rm sin}^2 \phi D_{b \mu} \longrightarrow  
U({\rm cos}^2 \phi D_{\mu} + {\rm sin}^2 \phi D_{b \mu} )U^{+}, 
\eqno{(8.1)}
$$
where $\phi$ is constant. If $D_{\mu}$ in eq(2.8) is replaced by ${\rm
cos}^2 \phi D_{\mu}  + {\rm sin}^2 \phi D_{b \mu}$, we can obtain the
following Lagrangian:
$$
\begin{array}{ccl}
\cal L &= &- \overline{\psi} \lbrack \gamma ^{\mu} ({\rm cos}^2 \phi D_{\mu} 
+{\rm sin}^2 \phi D_{b \mu}) +m \rbrack \psi  \\
&&-\frac{1}{4K} Tr( A^{\mu \nu} A_{\mu \nu} )
-\frac{1}{4K} Tr( B^{\mu \nu} B_{\mu \nu} ) \\
&&-\frac{\mu ^2}{2K ( 1+ \alpha ^2)} 
Tr \left \lbrack (A^{\mu}+\alpha B^{\mu})( A_{\mu}+\alpha B_{\mu} ) 
\right \rbrack
\end{array}
\eqno{(8.2)} 
$$
Obviously, this Lagrangian has local $SU(N)$ gauge symmetry. Let 
${\cal L}_{\psi}$ denote the part of the Lagrangian for fermions:
$$
{\cal L}_{\psi} = - \overline{\psi} 
\lbrack \gamma ^{\mu} ({\rm cos}^2 \phi D_{\mu} 
+{\rm sin}^2 \phi D_{b \mu}) +m \rbrack \psi.  
\eqno{(8.3)}
$$
Using eqs(2.4a,b), we can change ${\cal L}_{\psi}$ into 
$$
{\cal L}_{\psi} = - \overline{\psi} \lbrack \gamma ^{\mu} ( \partial _{\mu} 
- i g {\rm cos}^2 \phi A_{\mu} 
+ i \alpha g {\rm sin}^2 \phi B_{ \mu}) +m \rbrack \psi . 
\eqno{(8.4)}
$$
From the above Lagrangian, we know that both gauge fields $A_{\mu}$ and 
$B_{\mu}$ directly couple to matter field $\psi$. Substitute eqs(5.7a,b) 
into eq(8.4), we get
$$
{\cal L}_{\psi} = - \overline{\psi} \lbrack \gamma ^{\mu} ( \partial _{\mu} 
- i g  \frac{{\rm cos}^2 \theta -{\rm sin}^2 \phi}{{\rm cos }\theta}  C_{\mu} 
+ i g {\rm sin}\theta F_{ \mu}) +m \rbrack \psi . 
\eqno{(8.5)}
$$
\\

 The equation of motion for fermion field $\psi$ is
$$
\lbrack \gamma ^{\mu} ( \partial _{\mu} 
- i g  \frac{{\rm cos}^2 \theta -{\rm sin}^2 \phi}{{\rm cos }\theta}  C_{\mu} 
+ i g {\rm sin}\theta F_{ \mu}) +m \rbrack \psi = 0 . 
\eqno{(8.6)}
$$
The equations of motion for gauge fields $A_{\mu}$ and $B_{\mu}$  
now change into:
$$
D^{\mu} A_{\mu \nu}- \frac{\mu^2}{1+\alpha^2} ( A_{\nu} + \alpha B_{\nu}) = 
i g {\rm cos}^2 \phi  \overline{\psi} \gamma _{\nu} T^i \psi T^i
\eqno{(8.7a)}
$$
$$
D_b^{\mu} B_{\mu \nu}- \frac{\alpha \mu^2}{1+\alpha^2} 
( A_{\nu} + \alpha B_{\nu}) = 
- i \alpha g {\rm sin}^2 \phi  \overline{\psi} \gamma _{\nu} T^i \psi T^i .
\eqno{(8.7b)}
$$
\\

 If $\phi$ vanish, the Lagrangian density (8.2) will return to  the 
original Lagrangian density (2.8), the equations of motion (8.7a,b) will 
return to eqs(6.2a,b), and eq(8.6) will return to eq(6.1). So, the model 
discussed in the above chapters is just a special case of the model we 
discuss in this chapter. \\

\section{U(1) Case}

~~~~ If the symmetry of the model is U(1) group, we will obtain a 
U(1) gauge field model. We also use $A_{\mu}$ and $B_{\mu}$  to 
denote gauge fields and $\psi$ to denote a fermion 
fields. In U(1) case, the strengths of gauge fields are
$$
A_{\mu \nu}  =  \partial _{\mu} A_{\nu} - \partial _{\nu} A_{\mu}
\eqno{(9.1a)} 
$$
$$
B_{\mu \nu}  =  \partial _{\mu} B_{\nu} - \partial _{\nu} B_{\mu}
\eqno{(9.1b)} 
$$
Two gauge covariant derivatives are the same as (2.4a,b) but with 
different content. The Lagrangian density of the model is:
$$
\begin{array}{ccl}
\cal L &= &- \overline{\psi} 
\lbrack \gamma ^{\mu} ({\rm cos}^2 \phi D_{\mu} 
+{\rm sin}^2 \phi D_{b \mu}) +m \rbrack \psi  \\
&&-\frac{1}{4}  A^{\mu \nu} A_{\mu \nu} 
-\frac{1}{4}  B^{\mu \nu} B_{\mu \nu}  \\
&&-\frac{\mu ^2}{2 ( 1+ \alpha ^2)} 
 (A^{\mu}+\alpha B^{\mu})( A_{\mu}+\alpha B_{\mu} ) 
\end{array}
\eqno{(9.2)} 
$$
Local U(1) gauge transformations are
$$
\psi \longrightarrow e^{- i \theta} \psi ,
\eqno{(9.3a)} 
$$
$$
A_{\mu} \longrightarrow  A_{\mu} -\frac{1}{g} \partial _{\mu}\theta
\eqno{(9.3b)} 
$$
$$
B_{\mu} \longrightarrow  B_{\mu} +\frac{1}{ \alpha g} \partial _{\mu}\theta .
\eqno{(9.3c)} 
$$
Then, $A_{\mu \nu} ,~B_{\mu \nu}$ and $A_{\mu}+\alpha B_{\mu}$ are 
all U(1) gauge invariant.  That is
$$
A_{\mu \nu} \longrightarrow  A_{\mu \nu}
\eqno{(9.4a)} 
$$
$$
B_{\mu \nu} \longrightarrow  B_{\mu \nu}
\eqno{(9.4b)} 
$$
$$
A_{\mu}+\alpha B_{\mu} \longrightarrow A_{\mu}+\alpha B_{\mu}
\eqno{(9.4c)} 
$$
Using all these results, it is easy to prove that the Lagrangian 
density given by eq(9.2) has local $U(1)$ gauge symmetry.  \\

 Substitute eqs(5.7a,b) into eq(9.2), the Lagrangian density 
${\cal L}$ changes into
$$
\begin{array}{ccl}
{\cal L} & = & - \overline{\psi} \lbrack \gamma ^{\mu} ( \partial _{\mu} 
- i g  \frac{{\rm cos}^2 \theta -{\rm sin}^2 \phi}{{\rm cos }\theta}  C_{\mu} 
+ i g {\rm sin}\theta F_{ \mu}) +m \rbrack \psi \\ 
&&-\frac{1}{4}  C^{\mu \nu} C_{\mu \nu} 
-\frac{1}{4}  F^{\mu \nu} F_{\mu \nu} - \frac{\mu ^2}{2} C^{\mu} C_{\mu}
\end{array}
\eqno{(9.5)} 
$$
where,
$$
C_{ \mu \nu} = \partial _{\mu} C_{\nu} - \partial _{\nu} C_{\mu}
\eqno{(9.6a)} 
$$
$$
F_{ \mu \nu} = \partial _{\mu} F_{\nu} - \partial _{\nu} F_{\mu}
\eqno{(9.6b)} 
$$
So, in this model,  there is a massive Abel  gauge field as well 
as a massless Abel gauge field. They all have gauge interactions 
with matter field. In this case, $U(1)$ gauge interactions is 
transmitted by two different kinds of gauge fields.\\

It is known that QED is a U(1) gauge field theory. According to 
the model we discuss here, we may guess that there may exist two 
different kinds of photon, one is massive while another is massless. 
And if $\theta$ is near $\pi /2$ and the mass of massive photon is 
large, the massless photon field couples with charged fields 
will be much stronger than massive photon.   \\

\section{Two limits of the model.}

~~~~ Now, we let's discuss  two kinds of limits of this model. 
The first kind of limits corresponds to very small parameter $\alpha$. Let
$$
\alpha \longrightarrow 0 ,
\eqno{(10.1)}
$$
then
$$
{\rm cos} \theta \approx 1 ~~,~~ {\rm sin}\theta \approx 0.
\eqno{(10.2)}
$$
From eqs(5.6a,b), we know that the gauge field $A_{\mu}$ is just gauge 
field $C_{\mu}$ and the gauge field $B_{\mu}$ is just the gauge 
field $F_{\mu}$. That is
$$
C_{\mu} \approx A_{\mu} ~~,~~ F_{\mu} \approx B_{\mu} .
\eqno{(10.3)}
$$
In this case, the Lagrangian density (2.9) becomes
$$
\begin{array}{ccl}
{\cal L} & \approx & 
- \overline{\psi} \lbrack \gamma ^{\mu} ( \partial _{\mu} 
- i g  C^i_{\mu} T^i ) +m \rbrack \psi \\ 
&&-\frac{1}{4}  C^{i \mu \nu} C^i_{\mu \nu} 
-\frac{1}{4}  F^{i \mu \nu} F^i_{\mu \nu} 
- \frac{\mu ^2}{2} C^{i \mu} C^i_{\mu} .
\end{array}
\eqno{(10.4)} 
$$
The massless gauge field do not interact with matter fields
in this limit. So, the  $\alpha 
\longrightarrow 0 $ limit corresponds to the case that gauge 
interactions are mainly transmitted by 
massive gauge field. So, the above Lagrangian approximately 
describes those kinds of gauge interactions  
which are dominated by massive gauge bosons.   \\

The second kind of limits corresponds to very big parameter $\alpha$. Let
$$
\alpha \longrightarrow \infty ,
\eqno{(10.5)}
$$
then
$$
{\rm cos} \theta \approx 0 ~~,~~ {\rm sin}\theta \approx 1.
\eqno{(10.6)}
$$
From eqs(5.6a,b), we know that:
$$
C_{\mu} \approx B_{\mu} ~~,~~ F_{\mu} \approx - A_{\mu} .
\eqno{(10.7)}
$$
Then, the Lagrangian density (2.9) becomes
$$
\begin{array}{ccl}
{\cal L} & \approx & 
- \overline{\psi} \lbrack \gamma ^{\mu} ( \partial _{\mu} 
+ i g  F^i_{\mu} T^i ) +m \rbrack \psi \\ 
&&-\frac{1}{4}  F^{i \mu \nu} F^i_{\mu \nu} 
-\frac{1}{4}  C^{i \mu \nu} C^i_{\mu \nu} 
- \frac{\mu ^2}{2} C^{i \mu} C^i_{\mu} .
\end{array}
\eqno{(10.8)} 
$$
In this case, massive gauge field does not directly interact with matter 
fields. So, this limit corresponds to the case when gauge interactions are 
mainly transmitted by massless gauge field. \\

 In the particles' interaction model which describes the gauge
interactions of real world, the 
parameter $\alpha$ is finite,
$$
0 < \alpha < \infty.
\eqno{(10.9)}
$$
In this case, both massive gauge field and massless gauge field directly 
interact with matter fields, and 
gauge interactions are transmitted by both of them. \\

\section{The renormalizability of the theory}

~~~~ The renormalizability of the theory can
be very strictly proved\cite{15}. 
But this proof is extremely long and is not
suitable to write it here. We will not discuss it in details in 
this paper. We only want to discuss some key problems on the
renormalizability of the theory.  \\

It is know that, according to the power counting law, a massive
vector field model is not renormalizable in most case. The reason is simple.
It is known that the propagator of a massive vector field usually has the
following form:
$$
\Delta_{F \mu \nu} = 
\frac{-i}{k^2 + \mu^2 - i \varepsilon} 
(g_{\mu \nu} + \frac{k_{\mu} k_{\nu}}{\mu^2} ).
\eqno{(11.1)}
$$
So, when we let
$$
k \longrightarrow \infty
\eqno{(11.2)}
$$
then,
$$
\Delta_{F \mu \nu} 
\longrightarrow const .
\eqno{(11.3)}
$$
In this case, there are infinite kinds of divergent Feynman diagrams.
According to the power counting law, this theory is a kind of
non-renormalizable theory. Though  gauge field theory
contains massive vector fields, it is renormalizable. The key reason is that
the Lagrangian has local gauge symmetry\lbrack 8 \rbrack.  \\

 As we have stated before, this gauge field theory has
maximum local $SU(N)$ gauge symmetry. When we quantize this gauge
field theory in the path integral formulation, we must select gauge
conditions first \lbrack 17 \rbrack. In order to make gauge transformation
degree of freedom completely fixed, we must select two gauge conditions
simultaneously: one is for massive gauge field $C_{\mu}$ and another is for
massless gauge field $F_{\mu}$. For example, if we select temporal gauge
condition for massless gauge field $F_{\mu}$:
$$
F_4 = 0 ,
\eqno{(11.4)}
$$
there still exists remainder gauge transformation degree of freedom, because
temporal gauge condition is unchanged under the following local gauge
transformation:
$$
F_{\mu} \longrightarrow U F_{\mu} U^{\dag}
+\frac{1}{i g {\rm sin}\theta } U \partial _{\mu} U^{\dag}
\eqno{(11.5)}
$$
where
$$
\partial_t U = 0,  U = U(\stackrel{\rightarrow}{x}) .
\eqno{(11.6)}
$$
In order to make this remainder gauge transformation degree of freedom
completely fixed, we'd better select another gauge condition for gauge
field $C_{\mu}$. For example, we can select the following gauge condition
for gauge field $C_{\mu}$:
$$
\partial^{\mu} C_{\mu} = 0.
\eqno{(11.7)}
$$
\\

 If we select two gauge conditions simultaneously, when we quantize the
theory in path integral formulation, there will be two gauge fixing terms in
the effective Lagrangian. The effective Lagrangian can be written as:
$$
{\cal L}_{eff} = {\cal L} 
- \frac{1}{2 \alpha_1} f_1^a f_1^a
- \frac{1}{2 \alpha_2} f_2^a f_2^a
+ \overline{\eta}_1 M_{f1} \eta_1
+ \overline{\eta}_2 M_{f2} \eta_2
\eqno{(11.8)}
$$
where 
$$
f_1^a = f_1^a (F_{\mu}) ,
f_2^a = f_2^a (C_{\mu})
\eqno{(11.9)}
$$
If we select 
$$
f_2^a = \partial^{\mu} C_{\mu}^a ,
\eqno{(11.10)}
$$
then the propagator for massive gauge field $C_{\mu}$ is:
$$
\Delta_{F \mu \nu}^{ab} (k) =
\frac{-i \delta^{ab}}{k^2 + \mu^2 - i \varepsilon}
\left ( g_{\mu \nu} -(1- \frac{1}{\alpha_2}) 
\frac{k_{\mu} k_{\nu}}{k^2 - \mu^2/ \alpha_2} \right).
\eqno{(11.11)}
$$
If we let $k$ approach infinity, then
$$   
\Delta_{F \mu \nu}^{ab} (k) 
\sim \frac{1}{k^2} .
\eqno{(11.12)}
$$
In this case, according to the power counting law, the 
gauge field theory which is discussed in this paper is a 
kind of renormalizable theory. At the same
time, the local $SU(N)$ gauge symmetry will give a Ward-Takahashi identity
which will eventually make the theory renormalizable. A strict proof on the
renormalizability of the  gauge field theory can be found in the
reference \lbrack 15 \rbrack.  \\

 In order to make the  gauge field theory renormalizable, it is
very important to keep the maximum local $SU(N)$ gauge symmetry of the
Lagrangian. From the above discussions, we know that, in the
renormalization of the gauge field theory, local gauge symmetry
plays the following two important roles: 1) to make the propagator of the
massive gauge bosons have the renormalizable form; 2) to give a
Ward-Takahashi identity which plays a key role in the proof of the 
renormalizability of the gauge field theory.  \\

\section{Comments}

Up to now, we know that there are two mechanisms that can make
gauge field obtain non-zero mass: one is the Higgs mechanism which 
is well known in constructing the Standard Model; another is the 
mechanism discussed in this paper. In this new mechanism, the mass 
term of gauge field is introduced by using another set of gauge field
and the mass term of gauge fields does not affect the symmetry of 
the Lagrangian. We can imagine the new interaction picture as:
when matter fields take part in gauge interactions,  they emit or
absorb one kind of gauge field which is not eigenstate of 
mass matrix, when we detect this gauge fields in experiments, it will
appear in two states which corresponds to 
two kinds of vector fields, one is massless and another is 
massive.   \\

Though the Lagrangian of the model contains the mass term of gauge 
fields, the theory is renormalizable. So, we can use the mechanism
to describe gauge interactions of quarks and leptons. If we 
apply this mechanism to electroweak interactions, we can construct
an electroweak model which contains no Higgs particle. \\

\end{document}